\documentclass[prl,reprint,amsmath,amssymb,aps,preprintnumbers,showpacs]{revtex4-1}
\usepackage[colorlinks=true,linkcolor=black, citecolor=black,urlcolor=black]{hyperref} 
\usepackage{amstext,amsmath,amssymb,amsfonts,mathrsfs}   
\usepackage{graphicx}  
\usepackage{braket} 
\usepackage{bbm} 
\usepackage[british,UKenglish,USenglish,english,american]{babel}
\usepackage[dvipsnames]{xcolor}
 \usepackage{tikz}
 \usetikzlibrary{plotmarks,calc,decorations, decorations.pathmorphing, patterns,hobby}
 \usetikzlibrary{arrows}
 \tikzstyle dynkin node=[very thick,shape=circle,draw,inner sep=0pt,minimum size=5mm]
 \tikzstyle dynkin line=[very thick]
 \tikzstyle inverse line=[gray,line width=1.46pt,line cap=round, dash pattern=on 0pt off 2\pgflinewidth]
 \tikzstyle red phase=[red,decoration={snake,amplitude=0.1mm,segment length=1.6mm},decorate]
 \tikzstyle blue phase=[blue,decoration={snake,amplitude=0.1mm,segment length=0.9mm},decorate]
 \tikzstyle green phase=[green,decoration={snake,amplitude=0.1mm,segment length=0.9mm},decorate]
 \tikzstyle brown phase=[brown,decoration={snake,amplitude=0.1mm,segment length=0.9mm},decorate]
 
 \usetikzlibrary{decorations.pathmorphing}
 \tikzstyle arrow=[thick,rounded corners=18pt,-latex]
 \tikzstyle box=[draw,rounded corners,outer sep=4pt]
\tikzstyle B node=[outer sep=0pt]
\tikzstyle Q node=[inner sep=1pt,outer sep=0pt]

\definecolor{purple_nice}{rgb}{0.4,0.2,0.7}
\definecolor{fuel_blue}{RGB}{42,162,185}
\definecolor{YInMn_blue}{RGB}{46, 80, 144}
\definecolor{ultramarine}{RGB}{63, 0, 255}
\definecolor{KLEIN_blue}{rgb}{0, 0.18, 0.65}

\newcommand{\de}{\text{d}}

\newcommand{\TTbar}{T\overline{T}}

\begin{document}

\vspace*{0cm}

\title{$\TTbar$ deformations and integrable spin chains}
\author{Enrico Marchetto$^{1,2}$}
\email{enrico.marchetto.4@studenti.unipd.it}
\author{Alessandro Sfondrini$^{3}$}
\email{sfondria@itp.phys.ethz.ch}
\author{Zhou Yang$^{3}$}
\email{yangzho@student.ethz.ch}

\affiliation{${}^1$  Dipartimento di Fisica e Astronomia, Universit\`a degli Studi di Padova, via Marzolo 8, 35131 Padova, Italy\\
${}^2$  Scuola Galileiana di Studi Superiori, Universit\`a degli Studi di Padova, Via S.~Massimo 33, 35129 Padova, Italy\\
${}^3$ Institut f\"ur theoretische Physik, ETH Z\"urich, Wolfgang-Pauli-Stra{\ss}e 27, 8093 Z\"urich,
Switzerland}
\date{\today}

\begin{abstract}
\noindent
We consider current-current deformations that generalise $\TTbar$ ones, and show that they may be also introduced for integrable spin chains. In analogy with the integrable QFT setup, we define the deformation as a modification of the S matrix in the Bethe equations. Using results by Bargheer, Beisert and Loebbert we show that the deforming operator is composite and constructed out of two currents on the lattice; its expectation value factorises like for $\TTbar$. Such a deformation may be considered for any combination of charges that preserve the model's integrable structure.
\end{abstract}

\maketitle

\paragraph{Introduction.}
Exactly-solvable models play a crucial role in theoretical physics. Important examples arise in lattice systems, such as spin chains, and in two-dimensional quantum field theory (QFT). Integrable (quantum) spin chains are known since the pioneering work of Bethe~\cite{Bethe:1931hc}, who showed how to characterise the spectrum of the Heisenberg model in terms of a simple set of polynomial equations. To date, the technique to solve this and other  more complicated spin chains goes under the name of Bethe ansatz~\cite{Faddeev:1996iy}, see also ref.~\cite{Levkovich-Maslyuk:2016kfv} for a recent review.
Bethe ansatz methods found applications also in two-dimensional QFTs --- we talk then of integrable QFTs (IQFTs) --- even though the details there are more involved, as it may be expected. Regardless, their physics is similar: integrable spin chains as well as IQFTs possess an infinite number of conserved charges, mutually commuting, which greatly constrain their dynamics (see \textit{e.g.}\ refs.~\cite{Dorey:1996gd,Bombardelli:2016rwb} for reviews of integrability).

Integrable models are not easy to find. Therefore, it often makes sense to construct new models as \emph{deformations} of known ones. For two-dimensional QFTs, one such way to construct models is to consider current-current deformations, that is to say to define an $\alpha$-deformed Hamiltonian $\mathbf{H}(\alpha)$ by the differential equation 
\begin{equation}
\label{eq:JJ}
    \frac{\de}{\de \alpha}\mathbf{H}(\alpha)=
    \mathcal{O}_{\mathcal{X}\mathcal{Y}}=
    \int\de x\, \mathcal{X}^\mu(x;\alpha)\,\mathcal{Y}^\nu(x;\alpha)\,\epsilon_{\mu\nu}\,,
\end{equation}
where $\mathcal{X}^\mu$ and $\mathcal{Y}^\mu$ are conserved currents,
\begin{equation}
    \label{eq:continuity}
    \frac{\partial}{\partial t} \mathcal{X}^0= -\frac{\partial}{\partial x} \mathcal{X}^1\,,\qquad    \frac{\partial}{\partial t} \mathcal{Y}^0= -\frac{\partial}{\partial x} \mathcal{Y}^1\,.
\end{equation}
It can be shown following~\cite{Zamolodchikov:2004ce} that the composite operator~$\mathcal{O}_{\mathcal{X}\mathcal{Y}}$ is well defined at the quantum level by point-splitting regularisation. Moreover, its expectation value factorises on energy and momentum eigenstates,
\begin{equation}
\label{eq:JJfactorisation}
    \langle \mathcal{O}_{\mathcal{X}\mathcal{Y}}\rangle = 
    \int\de x\, \langle\mathcal{X}^\mu(\alpha)\rangle\,\langle\mathcal{Y}^\nu(\alpha)\rangle\,\epsilon_{\mu\nu}\,.
\end{equation}

One well-studied setup is when both currents arise from the same irrotational conserved current $\mathcal{J}^\mu$. In this case setting~$\mathcal{X}^\mu\equiv \mathcal{J}^\mu$ and $\mathcal{Y}^\mu\equiv \epsilon^{\mu\nu}\mathcal{J}_\nu$ gives rise to the so-called $J\bar{J}$ deformations. These are very natural for two-dimensional \emph{conformal} QFTs (CFTs), as they preserve scale invariance. The well-definedness of~$\mathcal{O}_{\mathcal{X}\mathcal{Y}}$ and the factorisation~\eqref{eq:JJfactorisation} are quite natural then, as we are dealing with chiral and antichiral currents.

More recently, another current-current deformation has attracted much attention: the $\TTbar$ deformation~\cite{Smirnov:2016lqw,Cavaglia:2016oda}. This can be defined as in~\eqref{eq:JJ} by setting $\mathcal{X}^\mu=T^{\mu0}$ and $\mathcal{Y}^\mu=T^{\mu1}$. The resulting theory is Poincar\'e invariant, and it has numerous intriguing properties. First of all, the factorisation~\eqref{eq:JJfactorisation} together with the relation of~$T^{\mu\nu}$ to energy and momentum allows us to turn~\eqref{eq:JJ} into a flow equation for the energy levels of the theory,
\begin{equation}
\label{eq:Burgers}
    \partial_\alpha H(R,\alpha) = H(R,\alpha)\partial_RH(R,\alpha) + \frac{1}{R}P{}^2\,.
\end{equation}
Here $H(R,\alpha)$ is the energy of a given state in volume~$R$ in the $\alpha$-deformed theory, while $P=2\pi n/R$ is the (quantised) momentum. This yields the spectrum of \emph{any} deformed theory from the undeformed one.
Moreover, this sort of deformation is very good at respecting symmetries, such as supersymmetry~\cite{Baggio:2018rpv,Chang:2018dge, Jiang:2019hux,Chang:2019kiu}, modular invariance~\cite{Aharony:2018bad}, and most remarkably \textit{integrability}~\cite{Smirnov:2016lqw,Cavaglia:2016oda}. This means that if the original theory is integrable --- it possesses infinitely-many conserved charges that constrain its dynamics --- the deformed theory is too. This also applies if the original theory is a CFT, by virtue of its integrable structure, \textit{cf.}~\cite{Bazhanov:1994ft,Bazhanov:1998dq,Bazhanov:1996dr}. Despite this constrained structure, our understanding of $\TTbar$ deformations is far from complete. As the deformation is \emph{irrelevant}, the resulting theory is quite unusual from a Wilsonian point of view. These deformations seem related to gravity~\cite{Dubovsky:2017cnj,Dubovsky:2018bmo,Conti:2018tca,Ishii:2019uwk}, random geometries~\cite{Cardy:2018sdv}, and string theory~\cite{Cavaglia:2016oda,Baggio:2018gct, Frolov:2019nrr,Hashimoto:2019wct, Sfondrini:2019smd,Callebaut:2019omt,Tolley:2019nmm} (see also \cite{Dubovsky:2012wk,Caselle:2013dra} for earlier observations of the relation between strings and $\TTbar$, and~\cite{McGough:2016lol,Giveon:2017nie,Chakraborty:2019mdf} for relations with holography).

To obtain more insight into such a deformation we may try to define it in the presumably simpler framework of quantum mechanics, as opposed to QFT. Work in this direction, inspired by holography, was done in~\cite{Gross:2019ach}. Here we take a different road, focussing on integrable models. Both IQFTs and integrable spin chains are described by Bethe ansatz techniques. Moreover, $\TTbar$ deformations may be \emph{defined} using the Bethe ansatz~\cite{Cavaglia:2016oda}. This is even true for generalised versions of $\TTbar$, where the currents in~\eqref{eq:JJ} are chosen among the infinitely-many conserved commuting charges of the theory, as suggested in~\cite{Smirnov:2016lqw} (one needs however to use generalisations of the Bethe ansatz machinery~\cite{Hernandez-Chifflet:2019sua}).

Below we introduce a spin-chain version of $\TTbar$ deformations (more generally, of current-current deformations) starting from the Bethe ansatz, which we briefly review. Our task is helped by previous studies of integrable spin-chain deformations~\cite{Bargheer:2008jt,Bargheer:2009xy}. It is then easy to see that the deforming operator obeys a discretised version of~\eqref{eq:JJ}, and that the spin-chain equivalent of~$\mathcal{O}_{\mathcal{X}\mathcal{Y}}$ also factorises like in~\eqref{eq:JJfactorisation}. Still the resulting deformations are all but trivial, as we discuss on some examples.

\paragraph{Note added.}
At a late stage of this work, Ref.~\cite{Pozsgay:2019ekd} appeared; therein, among other things, the relation between $\TTbar$ deformations and Refs.~\cite{Bargheer:2008jt,Bargheer:2009xy} was also noted and used to obtain a discretised version of~\eqref{eq:JJ}, which appears to agree with our own~\eqref{eq:discreteJJ-no-r}.

\paragraph{Integrable deformations in the Bethe Ansatz.}
When we consider an IQFT in large volume, that is with $R\gg 1/m$ if $m$ is the typical mass-scale of the theory, the spectrum can be approximately described by the Bethe-Yang equations. Schematically,
\begin{equation}
\label{eq:Bethe-Yang}
    e^{i p_jR}\prod_{k\neq j}^N S(p_j,p_k)=1\,,\quad j=1,\dots N\,,
\end{equation}
for a state containing $N$ particles of momenta $p_1, \dots p_N$. Eq.~\eqref{eq:Bethe-Yang} is valid for a theory with a single flavour, so that the S-matrix $S(p_1,p_2)$ is a $\mathbb{C}$-number; the case of many flavours can be addressed using nested Bethe equations, see \textit{e.g.}~\cite{Levkovich-Maslyuk:2016kfv}. For our discussion this is merely a technical complication. It is also worth stressing that the finite-$R$ spectrum is given by a set of thermodynamic Bethe ansatz (TBA), which differ from the Bethe-Yang results by terms of order~$e^{-mR}$~\cite{Luscher:1985dn,Luscher:1986pf}. We refrain from introducing them only because they would somewhat obscure our exposition. With these caveats in mind, \eqref{eq:Bethe-Yang} provides a set of quantisation conditions for the momenta~$p_j$. The energy can be then computed when all particles are well-separated, and is given by
\begin{equation}
\label{eq:energy}
    H= \sum_{j=1}^N H(p_j)\,,
\end{equation}
where for a relativistic theory $H(p)=\sqrt{p^2+m^2}$.
Let us now define a deformation of such a theory by means of a CDD factor~\cite{Castillejo:1955ed} constructed out of a symplectic form on the space of commuting conserved charges of the theory~\cite{Arutyunov:2004vx,Arutyunov:2009ga}. This modifies~\eqref{eq:Bethe-Yang} as
\begin{equation}
\label{eq:CDD}
    S(p_j,p_k) \to e^{i\alpha(X_j Y_k-X_k Y_j)}\,S(p_j,p_k)\,,
\end{equation}
where $X_j$ is the value of some charge~$\mathbf{X}$ on the state~$|p_j\rangle$, and similarly~$Y_j$.
Eqs.~(\ref{eq:Bethe-Yang},~\ref{eq:CDD}) may be rewritten in terms of the \emph{original} S-matrix as
\begin{equation}
\label{eq:deformation}
    e^{i p_jR+i\alpha(X_j Y-Y_j X)}\prod_{k\neq j}^N S(p_j,p_k)=1.
\end{equation}
Here $X$ and $Y$ are the \emph{total} values of the charges, \textit{cf.}~\eqref{eq:energy}. A $\TTbar$ deformation arises for $X_j=p_j$ and $Y_j=H(p_j)$. Taking for simplicity $P=0$, we immediately see that the total-energy contribution $\alpha H$ shifts the volume~$R$. The flow equation~\eqref{eq:Burgers} may be derived from this construction, or from the TBA construction (also for $P\neq0$)~\cite{Cavaglia:2016oda}. This setup is even more general: $X$ and $Y$ can be any two commuting charges acting diagonally on well-separated multi-particle states~$|p_1,\dots p_N\rangle$. \textit{E.g.}, they may be flavour charges, or they may belong to the infinite family of mutually-commuting charges of the IQFT. The general setup~(\ref{eq:CDD}--\ref{eq:deformation}) is our starting point for discussing spin-chain deformations.

\paragraph{Bethe ansatz for integrable spin-chains.}
Let us review the Bethe ansatz for spin chains. A good example to keep in mind is the Heisenberg model (see \textit{e.g.}~\cite{Faddeev:1996iy,Levkovich-Maslyuk:2016kfv}). However, much of what we will say applies more generally. A spin-chain is a one-dimensional model of $R$ ordered sites, each hosting a spin (the fundamental representation of $su(2)$ for the Heisenberg case). The spin-chain dynamics is defined by the \emph{nearest-neighbour} Hamiltonian
\begin{equation}
\mathbf{H} = \sum_j \mathbf{h}_{j,j+1}\,.
\end{equation}
In the Heisenberg case $\mathbf{h}_{j,j+1}$ is essentially the permutation operator which swaps two neighbouring spins. It turns out that like in IQFT, in infinite volume~$R=\infty$ $\mathbf{H}$ is just one of infinitely-many conserved charges, all mutually commuting. Here they can all be generated from appropriately expanding the ``transfer matrix'', see \textit{e.g.}~\cite{Faddeev:1996iy,Levkovich-Maslyuk:2016kfv}. Higher charges are less and less local. We may indicate them as $\mathbf{H}_n$, where $n$ both labels the charge and indicates its range (in this sense~$\mathbf{H}\equiv\mathbf{H}_2$). When $R$ is finite and we impose periodic boundary conditions, we may write down a set of Bethe ansatz equations~\cite{Bethe:1931hc,Faddeev:1996iy,Levkovich-Maslyuk:2016kfv}. Formally this reads exactly like~\eqref{eq:Bethe-Yang}. Now $p_j$ are momenta of some \emph{magnons} --- fictitious excitations over an arbitrary vacuum. In practice, the vacuum state is given by the single lowest-spin state in the Hilbert space (all spins ``pointing down''); this is not necessarily the lowest-energy state. A magnon of momentum~$p$ corresponds to overturning a single spin in a plane-wave configuration of definite momentum. It is worth remarking that, while this system is not relativistic, it enjoys translation invariance. In particular, the shift operator~$\mathbf{U}$ which moves each site to the right,~$j\mapsto(j+1)$,  commutes with all $\mathbf{H}_n$ (both when $R=\infty$ and for periodic boundary conditions); the spin chain is \emph{homogeneous}. On a multi-magnon state we have that
\begin{equation}
\label{eq:shift}
    \mathbf{U}\,|p_1,\dots,p_N\rangle
    =e^{-i(p_1+\cdots + p_N)}\,|p_1,\dots,p_N\rangle\,.
\end{equation}
The energy of a state is still given by~\eqref{eq:energy}, up to a constant shift due to the vacuum energy which we will disregard. Similar formulae hold for the $\mathbf{H}_n$s, in terms of densities~$H_{(n)}(p)$ which may be determined from the transfer matrix, see \textit{e.g.}~\cite{Beisert:2013voa} for a few examples. 
Importantly, the dispersion relation for the various charges is no longer relativistic but it takes a periodic form, \textit{e.g.}\ $H(p)\approx \sin^2p$. This periodicity is a signature of the lattice structure; indeed had we explicitly introduced a spacing between the lattice sites this would have featured in the dispersion.

\paragraph{CDD deformations of a spin chain.}
Given that formally both IQFTs and (certain) integrable spin chains are described by Bethe equations, it is tempting to try to generalise $\TTbar$ deformations to spin chains through the CDD deformation~(\ref{eq:CDD}--\ref{eq:deformation}).
A natural question is what the deformed Hamiltonian $\mathbf{H}(\alpha)$ might be and whether it describes a \textit{bona-fide} spin-chain. Thankfully, this question was answered in broad generality in Refs.~\cite{Bargheer:2008jt,Bargheer:2009xy}. There the authors consider all deformations that preserve integrability and give rise to a local homogenous spin-chain order-by-order in~$\alpha$. In practice, they consider deformations induced by
\begin{equation}
\label{eq:diffeqdefo}
    \frac{\de}{\de \alpha} \mathbf{H}_n(\alpha) = 
    i\big[\mathbf{O}(\alpha),\mathbf{H}_n(\alpha)\big]\,,
\end{equation}
where $\mathbf{O}(\alpha)$ should be judiciously chosen so that the right-hand side is a local, homogeneous charge. The initial conditions are given by the undeformed~$\mathbf{H}_n$s. The upshot of this definition~\cite{Bargheer:2009xy} is that by Jacobi identity
\begin{equation}
    \frac{\de}{\de \alpha} \big[\mathbf{H}_n(\alpha),\mathbf{H}_m(\alpha)\big]=i\Big[\mathbf{O}(\alpha),\big[\mathbf{H}_n(\alpha),\mathbf{H}_m(\alpha)\big]
    \Big]\,.
\end{equation}
Hence the original algebra is preserved by such a deformation (in particular when $[\mathbf{H}_n,\mathbf{H}_m]=0$). This allows us to formally define a deformed transfer matrix by summing up the charges~$\mathbf{H}_n(\alpha)$. Moreover, \eqref{eq:diffeqdefo} may be formally integrated
\begin{equation}
\label{eq:Hint}
    \mathbf{H}_n(\alpha)=\mathbf{H}_n(0) + i\int\limits_0^\alpha\de\alpha'\big[\mathbf{O}(\alpha'),\mathbf{H}_n(\alpha')\big]\,,
\end{equation}
and solved perturbatively in small-$\alpha$~\cite{Bargheer:2009xy}, yielding longer-range terms at each order.
The authors of~\cite{Bargheer:2008jt,Bargheer:2009xy} list several choices of~$\mathbf{O}(\alpha)$. Remarkably, there is one that corresponds to (\ref{eq:CDD}--\ref{eq:deformation}), and is given by a \emph{bilocal} operator
\begin{equation}
\label{eq:bilocal}
    \mathbf{O}=[\mathbf{X}|\mathbf{Y}]= 
    \sum_{a\lesssim b} \mathbf{x}_{a}\,\mathbf{y}_b= 
    \sum_{a\lesssim b} \mathbf{y}_b\,\mathbf{x}_{a}\,,
\end{equation}
defined on an infinite chain.
This features the two local-operator densities $\mathbf{x}_{a},\mathbf{y}_{b}$ acting at sites $a,b$ with finite range. The sum is such that the two densities \emph{do not overlap}, and hence commute~%
\footnote{Note that our definition is slightly different from that of Bargheer, Beisert and Loebbert; however as they themselves note, the difference is a local charge, which would only yield a similarity transformation.}.
Under this condition, it can be shown~\cite{Bargheer:2009xy} that~\eqref{eq:bilocal} generates precisely the CDD factor~\eqref{eq:CDD} when used in eq.~\eqref{eq:diffeqdefo}, with $X_j$ in~\eqref{eq:CDD} satisfying $\mathbf{X}|p_j\rangle= X_j|p_j\rangle$ (and similarly $Y_j$).
Remark that this construction is rigorous for \emph{infinite} chains. The $\alpha$-expansion gives increasing-range operators, so that for finite chain the construction is correct up to ``wrapping order''~\cite{Bargheer:2009xy}.

\paragraph{The deforming operator.}
To make contact with ordinary $\TTbar$ deformations, let us work out explicit form of the flow equation~\eqref{eq:diffeqdefo} for the Hamiltonian (\textit{i.e.}, for $n=2$).
The right-hand side depends on $i[\mathbf{H},\mathbf{x}_a]=\tfrac{\de}{\de t}\mathbf{x}_a$ and $\tfrac{\de}{\de t}\mathbf{y}_b$ (recall that $\mathbf{H}\equiv\mathbf{H}_2$ is the generator of time-evolution),
\begin{equation}
\label{eq:Hflow}
    \frac{\de}{\de\alpha} \mathbf{H}(\alpha) = 
    -\sum_{a\lesssim b} \Big(
    \mathbf{x}_a\frac{\de \mathbf{y}_b}{\de t}+
    \mathbf{y}_b\frac{\de \mathbf{x}_a}{\de t}
    \Big)\,.
\end{equation}
Recall that $\mathbf{x}_a$ and  $\mathbf{y}_b$ commute~\eqref{eq:bilocal}.
To evaluate this expression, it is convenient to introduce a discretised version of the continuity equation~\eqref{eq:continuity},
\begin{equation}
\label{eq:discretediff}
    \frac{\de \mathbf{x}_a}{\de t} = - \Delta \chi_a \equiv \chi_{a-1}-\chi_a\,.
\end{equation}
where $\chi_a$ is the current corresponding to $\mathbf{x}_a$ at site~$a$; similarly $\mathbf{y}_a$ and $\eta_a$. Plugging~\eqref{eq:discretediff} into~\eqref{eq:Hflow}, one of the two sums telescopes and we find
\begin{equation}
\label{eq:discreteJJ}
    \frac{\de\mathbf{H}}{\de\alpha}  = \sum_a\mathcal{O}_{xy}(a,r)\equiv\sum_a\Big(
    \mathbf{y}_{a+r}\chi_a-\eta_{a+r-1}\mathbf{x}_{a}\Big)\,.
\end{equation}
To obtain this form, we assumed that the $\mathbf{x}_a$ and $\mathbf{y}_b$ were separated by a range~$r$ in~\eqref{eq:bilocal} and that we may discard the current flow at the end on the (infinite) chain.
This equation is reminiscent of the  ``current-current'' deformation~\eqref{eq:JJ}; it would match it if we could define a Lorentz vector~$\mathcal{X}^\mu(a)=(\mathbf{x}_a,\chi_a)$ --- a current with ``discrete'' conservation law~\eqref{eq:discretediff}. Even though this is improper, as the theory is not Lorentz invariant, it is worth noting that the expression in~\eqref{eq:discreteJJ} satisfies all the nice properties of current-current operators pointed out in~\cite{Zamolodchikov:2004ce}.
The expectation value $\langle\psi|\mathcal{O}_{xy}(a,r)|\psi\rangle$ on an eigenstate of $\mathbf{H}$ \emph{does not depend on $r$}; more precisely
\begin{equation}
\label{eq:Oisconstant}
    \Delta_{(r)} \langle\psi|\mathcal{O}_{xy}(a,r)|\psi\rangle=0\,.
\end{equation}
Eq.~\eqref{eq:Oisconstant} follows almost \textit{verbatim} from Zamolodchikov's arguments~\cite{Zamolodchikov:2004ce} up to trading the space-derivative for its discrete version~$\Delta$, and using the fact that the spin-chain is homogeneous. Therefore we can set~$r=0$ in~\eqref{eq:discreteJJ},
\begin{equation}
\label{eq:discreteJJ-no-r}
    \frac{\de}{\de\alpha} \langle\psi|\mathbf{H}(\alpha)|\psi\rangle = \sum_a\big\langle\psi\big|
    \mathbf{y}_{a}\chi_a-\eta_{a-1}\mathbf{x}_{a}\big|\psi\big\rangle\,,
\end{equation}
which closely reminds~\eqref{eq:JJ}~\footnote{%
We may think that we took a coincident-points limit like in Zamolodchikov's prescription for~$\TTbar$.
}.
Additionally, by $r$-independence and the fact that the spin-chain is homogenous, one can show that the expectation value of $\mathcal{O}_{xy}(a,r)$ factorises and
\begin{equation}
\label{eq:factorisation}
    \frac{\langle\psi|\mathcal{O}_{xy}|\psi\rangle}{R}=
    \langle\psi|\mathbf{y}|\psi\rangle\langle\psi|\chi|\psi\rangle-\langle\psi|\eta|\psi\rangle\langle\psi|\mathbf{x}|\psi\rangle\,,
\end{equation}
where we suppressed the dependence on~$a$ in $\mathbf{x}_a$, \textit{etc.}, thanks to translation invariance.
This holds on the eigenstates~$|\psi\rangle$ of $\mathbf{H}$ (and of $\mathbf{X}$ and $\mathbf{Y}$, which we assumed to commute with it). Eq.~\eqref{eq:factorisation} also holds on any state of the position-space basis, because the two operators do not act on the same sites.
%

\paragraph{Flow of higher charges.}

The machinery of Ref.~\cite{Bargheer:2009xy} allows us to study the variations of any of the mutually commuting charges $\mathbf{H}_n$, not just of $\mathbf{H}$. The main difference is that in deriving~\eqref{eq:Hflow} we will encounter commutators like~$i[\mathbf{H}_n,\mathbf{x}_a]$. These may be interpreted as the time-evolution generated from the ``Hamiltonian'' $\mathbf{H}_n$, similarly to the ``Hamiltonians'' which generate the invariant tori for the Liouville-Arnol'd theorem in classical mechanics (\textit{cf.}~\cite{arnold1997mathematical}). It was put forward in Ref.~\cite{Pozsgay:2019ekd} based on earlier work~\cite{Borsi:2019tim,Pozsgay:2019xak} that generalised current operators may be introduced with respect to these flows (the relation between such currents and long-range chains was noted in~\cite{Pozsgay:2019xak}). Once that is observed, all steps leading to Eq.~\eqref{eq:discreteJJ-no-r} may be repeated \textit{verbatim}, and all properties of $\mathcal{O}_{xy}(a,r)$ will hold by the same arguments.

\paragraph{Deformation by spin and energy.}

As a first example, we consider the Heisenberg chain and take the two operators appearing in~\eqref{eq:bilocal} to be the $su(2)$ spin along one preferred direction, and the Hamiltonian density, respectively:
\begin{equation}
    \mathbf{x}_a= \mathbf{s}_a\,,
    \qquad
    \mathbf{y}_b= \mathbf{h}_{b,b+1}\,.
\end{equation}
The spin~$\mathbf{S}$ commutes with~$\mathbf{H}$ and all other~$\mathbf{H}_n$.
When integrating~\eqref{eq:Hint} we encounter a simplification:
\begin{equation}
\label{eq:Sdoesnotflow}
    \frac{\de}{\de \alpha}\mathbf{S}=0\,.
\end{equation}
Still $\tfrac{\de}{\de \alpha}\mathbf{H}\neq0$.
The effect of such a deformation in the Bethe ansatz is to introduce a CDD factor of the form $(s_j H_k-s_kH_j)$. Up to a normalisation, $H_j \approx \sin^2(p_j)$; as for~$s_j$, each magnon increases the spin by one unit, $s_j=+1$. Therefore we get the (asymptotic) Bethe equations
\begin{equation}
    e^{ip_jR +i\alpha (H_j N-H)}\prod_{k\neq j}^N S(p_j,p_k)=1\,,
\end{equation}
while~\eqref{eq:energy} remains unchanged.
One immediate consequence, obvious from the form of our deformation, is that $su(2)$ invariance is broken~\footnote{This can also be seen from observing that symmetry descendant cannot be realised in the Bethe equations.}.
Such deformations were studied in~\cite{Beisert:2013voa} for the XXZ model, which has no~$su(2)$ symmetry from the get-go, but may be considered for any model with flavour symmetry at the price of breaking it to its Cartan subalgebra.

\paragraph{Deformations by higher charges.}
Refs.~\cite{Bargheer:2008jt,Bargheer:2009xy} focus on deformations by arbitrary combinations of the higher charges, \textit{i.e.}
\begin{equation}
    \mathbf{O} = [\mathbf{H}_n|\mathbf{H}_m]\,,\qquad
    n,m\geq 2\,.
\end{equation}
In principle these arbitrary deformations are defined in terms of the same ``current-current'' operator~\eqref{eq:discreteJJ-no-r}. In practice we expect the deformation to be fairly unwieldy already at first orders in~$\alpha$. It is worth remarking that even in QFT analogous deformations are only partially understood. In particular, recent efforts to describe generalised $\TTbar$ deformations involving higher charges have pointed to the necessity of introducing a ``mirror'' (in the sense of Refs.~\cite{Ambjorn:2005wa,Arutyunov:2007tc}) generalised Gibbs ensemble~\cite{PhysRevLett.98.050405}. It appears that deforming even a simple theory (say, a free CFT) in this fashion leads to intricate models.
It might be easier to consider the case of \textit{e.g.}\ $[\mathbf{S}|\mathbf{H}_n]$, as again~\eqref{eq:Sdoesnotflow} will hold.

\paragraph{Deformations by momentum and energy.}
Finally we come to the case which should most closely correspond to $\TTbar$ deformations, \textit{i.e.}\ that of~$[\mathbf{P}|\mathbf{H}]$. This immediately appears  problematic. Firstly, $\mathbf{P}$ is not a symmetry generator of the theory. Indeed, only \emph{finite} shifts (as opposed to \emph{infinitesimal} ones) may be realised on a spin chain. Momentum is related to the logarithm of the shift operator~$\mathbf{U}$, \textit{cf.}~\eqref{eq:shift}. Defining such an operator would require picking a branch. This is nicely illustrated by the would-be deformation of the Bethe equations,
\begin{equation}
\label{eq:TTbarbethe}
        e^{ip_j(R+\alpha H)-i\alpha H_j P}\prod_{k\neq j}^N S(p_j,p_k)=1\,.
\end{equation}
Even if we decided to restrict for simplicity to cyclically invariant chains for which $P=0$, we are still faced with a problem. These ``Bethe equations'' are not invariant under a periodic shift $p\to p+2\pi$, because $\alpha H$ (unlike~$R$) is not quantised. As a result these equations do not define a lattice system in the usual sense, even for small deformations.
It might be worth studying this problem in a ``covering space'' of sorts, to resolve the branches of the logarithm, bearing in mind that the construction of ref.~\cite{Bargheer:2009xy} is not directly applicable here because  $[\mathbf{P}|\mathbf{H}]$ is not bilocal.
CDD factors leading to equations like~\eqref{eq:TTbarbethe} do appear in many interesting models, chiefly in the $AdS_3/CFT_2$ correspondence~\cite{Borsato:2012ud,Beccaria:2012kb,Borsato:2013qpa,Borsato:2013hoa}, see ref.~\cite{Sfondrini:2014via} for a review. In that case there are even explicit examples of integrable string backgrounds whose finite-volume spectrum is described by the Bethe-Yang equations \emph{exactly}~\cite{Baggio:2018gct,Dei:2018mfl,Dei:2018jyj} --- the TBA trivialises. Because of the absence of wrapping effects and in view of the simplicity of the Bethe-Yang equations, these systems call for a quantum-mechanical (as opposed to QFT) interpretation, which the present framework might provide.

\paragraph{Conclusions and outlook.}
We have seen that, exploiting their Bethe-ansatz formulation, current-current deformations akin to~$\TTbar$ may be defined for spin chains in the framework developed by Bargheer, Beisert and Loebbert~\cite{Bargheer:2008jt,Bargheer:2009xy}. The discretised current-current composite operator~\eqref{eq:discreteJJ-no-r} satisfies the same properties of Zamolodchikov's $\TTbar$~\cite{Zamolodchikov:2004ce}. This points at the possibility of studying these types of deformations on one-dimensional lattices.

These deformations are well-defined for infinitely long chains~\cite{Bargheer:2008jt,Bargheer:2009xy} (otherwise, the deformation would ``wrap'' the chain). It is perhaps most interesting to exploit this framework with a continuum limit in mind. This looks like an interesting but highly non-trivial challenge. In that limit it should also be possible to recover the momentum operator, which so far does not seem to find a place in this discretised models. It would be interesting to understand the role of momentum better due to its importance for $AdS_3/CFT_2$ integrability~\cite{Borsato:2012ud,Beccaria:2012kb,Borsato:2013qpa,Borsato:2013hoa,Sfondrini:2014via,Baggio:2018gct,Dei:2018mfl,Dei:2018jyj}.
It is more straightforward conceptually, though perhaps cumbersome, to work with the higher charges~$\mathbf{H}_n$; these are the counterparts of the higher-spin charges in IQFTs, which can also be studied by generalised Bethe ansatz techniques~\cite{Hernandez-Chifflet:2019sua}. A major advantage of the deformation~\eqref{eq:discreteJJ-no-r}, together with its factorisation property, is the possibility at least in principle of writing flow equations similar to~\eqref{eq:Burgers} and to the generalised ones discussed in~\cite{LeFloch:2019wlf, Hernandez-Chifflet:2019sua}. It would be interesting to study these issue further and relate the IQFT and spin-chain pictures.

\begin{acknowledgments}
\paragraph{Acknowledgments.}
We are grateful to Niklas Beisert and Florian Loebbert for enlightening discussions on their older work which is at the heart of this paper. We also thank Andrea Dei, Christian Ferko, Sergey Frolov, Stefano Negro and Roberto Tateo for discussions and feedback on a draft of this manuscript. We thank Bal{\'a}sz Pozsgay for pointing out to us some related earlier work of his.
A.S.\ is grateful to Jorrit Kruthoff for discussions related to these ideas. He would also like to thank the participants of the workshop \textit{New frontiers of integrable deformations} for stimulating related discussions.
 A.S.'s work is funded by ETH Career Seed Grant No.~SEED-23 19-1, as well as by the NCCR SwissMAP, funded by the Swiss National Science Foundation.
\end{acknowledgments}

%

\end{document}